\documentclass[utf8]{FrontiersinHarvard} 

\usepackage{url,hyperref,lineno,microtype,subcaption}
\usepackage[onehalfspacing]{setspace}
\usepackage{amsmath}

\newcommand{\aap}{    {\it Astron. Astrophys.}}

\newcommand{\apj}{    {\it Astrophys. J.}}
\newcommand{\apjl}{   {\it Astrophys. J. Lett.}}

\newcommand{\mnras}{  {\it Mon. Not. Roy. Astron. Soc.}}

\newcommand{\solphys}{{\it Solar Phys.}}
 
\newcommand{\ssr}{    {\it Space Sci. Rev.}} 
\newcommand{\araa}{    {\it Ann. Rev. Astron. Astrophys.}} 
\newcommand{\apjs}{    {\it Astrophys. J. Suppl. Ser.}} 

\def\keyFont{\fontsize{8}{11}\helveticabold }
\def\firstAuthorLast{Kolotkov} 
\def\Authors{Dmitrii Y. Kolotkov\,$^{1,*}$}
\def\Address{$^{1}$Centre for Fusion, Space and Astrophysics, Physics Department, University of Warwick, Coventry CV4 7AL, United Kingdom}
\def\corrAuthor{Dmitrii Y. Kolotkov}

\def\corrEmail{D.Kolotkov.1@warwick.ac.uk}

\begin{document}
\onecolumn
\firstpage{1}

\title[Non-adiabatic coronal seismology]{Coronal seismology by slow waves in non-adiabatic conditions} 

\author[\firstAuthorLast ]{\Authors} 
\address{} 
\correspondance{} 

\extraAuth{}

\maketitle

\begin{abstract}
Slow magnetoacoustic waves represent an important tool for probing the solar coronal plasma. The majority of seismological methods with slow waves are based on a weakly non-adiabatic approach, which assumes the coronal energy transport has only weak effects on the wave dynamics. Despite it significantly simplifies the application of coronal seismology by slow waves, this assumption omits a number of important and confidently observed effects and thus puts strong limitations on the reliability of seismological estimations.
We quantitatively assess the applicability of the {weak thermal conduction} theory to coronal seismology by slow waves. We numerically model the linear standing slow wave in a 1D coronal loop, with field-aligned thermal conduction $\kappa_\parallel$ as a free parameter and no restrictions on its efficiency. The time variations of the perturbed plasma parameters, obtained numerically with full {conductivity}, are treated as potential observables and analysed with the standard data processing techniques.
The slow wave oscillation period is found to increase with $\kappa_\parallel$ by about 30\%, indicating the corresponding modification in the effective wave speed, which is missing from the {weak conduction theory}. Phase shifts between plasma temperature and density perturbations are found to be well consistent with the approximate {weakly conductive solution} for all considered values of $\kappa_\parallel$. In contrast, the comparison of the numerically obtained ratio of temperature and density perturbation amplitudes with the weak theory revealed relative errors up to 30--40\%. We use these parameters to measure the effective adiabatic index of the coronal plasma directly as the ratio of the effective slow wave speed to the standard sound speed and in the polytropic assumption, which is found to be justified in a {weakly conductive regime} only, with relative errors up to 14\% otherwise. The damping of the initial perturbation is found to be of a non-exponential form during the first cycle of oscillation, which could be considered as an indirect signature of entropy waves in the corona, also not described by {weak conduction theory}. The performed analysis and obtained results offer a more robust scheme of coronal seismology by slow waves, with reasonable simplifications and without the loss of accuracy.

\tiny
 \keyFont{ \section{Keywords:} Sun, Corona, Magnetohydrodynamics, Waves, Coronal Seismology} 
\end{abstract}

\section{Introduction}

The outermost layer of the solar atmosphere, the corona, consists of a fully ionised and strongly magnetised plasma, which is able to respond periodically or quasi-periodically to any impulsive perturbation. The interest in studying coronal waves and oscillations is connected with their possible role in the enigmatic coronal heating problem \citep{2020SSRv..216..140V} and with a promising and sometimes unique opportunity to probe the coronal plasma parameters with the method of magnetohydrodynamic (MHD) seismology \citep{2020ARA&A..58..441N}. In particular, fast magnetoacoustic wave modes, directly observed in the corona as e.g. kink oscillations of coronal loops or fast-propagating quasi-periodic wave trains, are extensively used for probing the coronal magnetic field strength and twist, density stratification, and cross-field fine structuring \citep[see e.g.][for the most recent comprehensive reviews]{2021SSRv..217...73N, 2020SSRv..216..136L, 2022SoPh..297...20S}. The slow mode of magnetoacoustic waves, which appear in standing \citep[e.g.][]{2021SSRv..217...34W}, propagating \citep[e.g.][]{2021SSRv..217...76B}, and sloshing \citep[e.g.][]{2019ApJ...874L...1N} forms, has in turn been found sensitive to both the magnetic and thermodynamic properties of the coronal plasma, which makes it a powerful seismological tool too.

The seismological applications of slow waves in the corona span from probing the absolute value of the magnetic field in active regions \citep[e.g.][]{2007ApJ...656..598W, 2016NatPh..12..179J} and the magnetic field direction \citep{2009ApJ...697.1674M} to estimating the effective adiabatic index of the coronal plasma \citep[e.g.][]{2011ApJ...727L..32V, 2015ApJ...811L..13W, 2018ApJ...868..149K}, its effective energy transport coefficients \citep{2015ApJ...811L..13W, 2018ApJ...860..107W}, multi-thermal nature of coronal loops \citep[e.g.][]{2003A&A...404L...1K, 2017ApJ...834..103K}, and even properties of the coronal heating function \citep{2019ApJ...884..131R, 2020A&A...644A..33K}. Moreover, a similarity between the properties of the phenomenon of quasi-periodic pulsations, observed in solar and stellar flare lightcurves and attributed to the modulation of the flare emission by slow waves, allowed for revealing new solar-stellar analogies \citep{2016ApJ...830..110C} and stimulated the development of the theory of slow waves in stellar coronal conditions \citep[e.g.][]{2018ApJ...856...51R, 2022ApJ...931...63L}.

The majority of seismological estimations with slow waves have been carried out under the assumption of weak non-adiabaticity of the coronal plasma, i.e. assuming the energy exchange and energy transfer processes (such as thermal conduction, compressive viscosity, optically thin radiation, etc) are weak and slow in comparison with the oscillation period of a slow wave as its characteristic timescale. Under this assumption, the seismological analysis with slow waves gets substantially simplified. However, it cannot properly account for such important observable effects as rapid damping of slow waves, with the damping time being about the oscillation period \citep[see e.g.][for the most recent multi-instrumental statistical survey]{2019ApJ...874L...1N}, apparently linear scaling between the slow wave damping time and oscillation period \citep[see e.g.][]{2016ApJ...830..110C, 2016ApJ...820...13M}, strong modification of a slow wave speed and effective adiabatic index of the corona \citep[see e.g.][who detected the effective adiabatic index to vary from about 5/3 to 1]{2018ApJ...868..149K}, large phase shifts between the plasma temperature and density perturbed by slow waves \citep{2019MNRAS.483.5499K}, and yet undetected effects such as coupling of the slow and entropy wave modes \citep{2021SoPh..296...96Z}. Furthermore, the transport coefficients of those non-adiabatic processes are often considered as free parameters in the corona, and their deviation from the classical values prescribed by \citet{1962pfig.book.....S} and \citet{1965RvPP....1..205B} due to essentially dynamic and turbulent nature of the coronal plasma remains a subject to intensive studies. In particular, the parametric study of the dynamics of slow waves in coronal loops with suppressed field-aligned thermal conduction and of their diagnostic potential was performed recently by \citet{2019ApJ...886....2W}. Likewise, the question of \lq\lq anomalous transport\rq\rq\ remains open in other astrophysical plasma environments too \citep[see e.g.][for the discussion of this topic in the Earth's magnetospheric plasma context]{2017PhPl...24b2104M}.

In this work, we delineate the applicability of a {weak thermal conduction theory} of slow waves to coronal seismology. For this, we numerically model the linear evolution of a standing slow oscillation in a hot coronal loop (alike those observed with the SUMER instrument onboard the SOHO spacecraft or in \lq\lq hot\rq\rq\ channels of SDO/AIA) with {full conductivity}, and compare the model outcomes to those obtained in a {weakly conductive limit}. In particular, we focus on the measurements of the phase shift and relative amplitude ratio between density and temperature perturbations and their use for probing the effective adiabatic index of the coronal plasma. The applicability of a polytropic assumption for estimating the effective adiabatic index is also discussed. The paper is structured as follows. In Sec.~\ref{sec:model}, we describe the numerical model and plasma loop parameters. In Sec.~\ref{sec:analysis}, we present the analysis of oscillatory variations of plasma loop density and temperature, caused by the standing slow wave, and the comparison of those in the numerical solution with {full conductivity} and in an approximate {weakly conductive limit}. The application of the obtained oscillation parameters to probing the effective adiabatic index of the coronal plasma, in the polytropic assumption and as ratio of the effective wave speed to the standard sound speed, is demonstrated in Sec.~\ref{sec:gamma}. The discussion of the obtained results and conclusions are summarised in Sec.~\ref{sec:disc}.

\section{Governing equations and modelling}
\label{sec:model}

We model the dynamics of a standing slow wave in a low-beta coronal plasma in the infinite magnetic field approximation \citep[see Sec. 2.3 of][and references therein]{2021SSRv..217...34W}, using the following set of linearised governing equations,
\begin{align}\label{eq_motion}
&\rho_0 \frac{\partial V_{z}}{\partial t} = -\frac{\partial p}{\partial z},\\
&\frac{\partial \rho}{\partial t} +  \rho_0\frac{\partial V_{z}}{\partial z} = 0,\\
&p=\frac{k_\mathrm{B}}{m}\left( \rho_0 T + T_0 \rho\right),\\
&\frac{\partial {T}}{\partial t} - (\gamma-1)\frac{T_0}{\rho_0}\frac{\partial {\rho}}{\partial t}=\frac{\kappa_\parallel}{\rho_0 C_\mathrm{V}}\frac{\partial^2{T}}{\partial z^2}.\label{eq:energy}
\end{align}
In Eqs.~(\ref{eq_motion})--(\ref{eq:energy}), the direction of the wave propagation along the $z$-axis is prescribed by the infinitely stiff (not perturbed by the wave) magnetic field; $V_z$, $p$, $\rho$, and $T$ represent perturbations of the plasma velocity, pressure, density, and temperature, respectively; the subscripts \lq\lq 0\rq\rq\ correspond to the values of those variables at $t=0$; $m$, $\gamma$, and $k_\mathrm{B}$ are the mean particle mass, standard adiabatic index $5/3$, and Boltzmann constant, respectively; $C_\mathrm{V} = (\gamma-1)^{-1}k_\mathrm{B}/m$ is the standard specific heat capacity; and the coefficient of thermal conduction along the field $\kappa_\parallel$ is treated as a free parameter in this study. The effects of other non-ideal processes, such as compressive viscosity, optically thin radiation, heating, and the wave-induced misbalance between them, are omitted, with the field-aligned thermal conduction considered as the main wave damping mechanism \citep[e.g.][]{2002ApJ...580L..85O, 2003A&A...408..755D, 2004A&A...415..705D, 2016ApJ...826L..20R, 2019A&A...628A.133K}.

The presence of a dissipative term on the right-hand side of energy equation (\ref{eq:energy}) makes the model essentially non-adiabatic and {may lead} to the appearance of the phase shift $\Delta \varphi$ between the temperature and density perturbations and modify the ratio of their instantaneous relative amplitudes $A_T$ and $A_\rho$, respectively \citep[see e.g. a series of works by][]{2009A&A...494..339O, 2011ApJ...727L..32V, 2015ApJ...811L..13W, 2018ApJ...868..149K, 2022SoPh..297....5P}. In a {weakly conductive limit}, i.e. assuming the damping time by thermal conduction is much longer than the oscillation period and the wave speed remains equal to the standard sound speed $C_\mathrm{s} = \sqrt{\gamma k_\mathrm{B}T_0/m}$, the parameters $\Delta \varphi$ and $A_T/A_\rho$ have been previously derived as
\begin{align}
	&\tan \Delta \varphi \approx \frac{2\pi(\gamma-1)\kappa_\mathrm\parallel m}{k_\mathrm{B}C_\mathrm{s}^2 P_0 \rho_0},\label{eq:phi_weak}\\
	&\frac{A_T}{A_\rho} \approx (\gamma-1)\cos \Delta \varphi,\label{eq:amp_weak}
\end{align}
where $P_0=2L/C_\mathrm{s}$ is oscillation period in the ideal adiabatic case, with $L$ being the loop length.

\begin{figure}
	\begin{center}
		\includegraphics[width=0.39\textwidth]{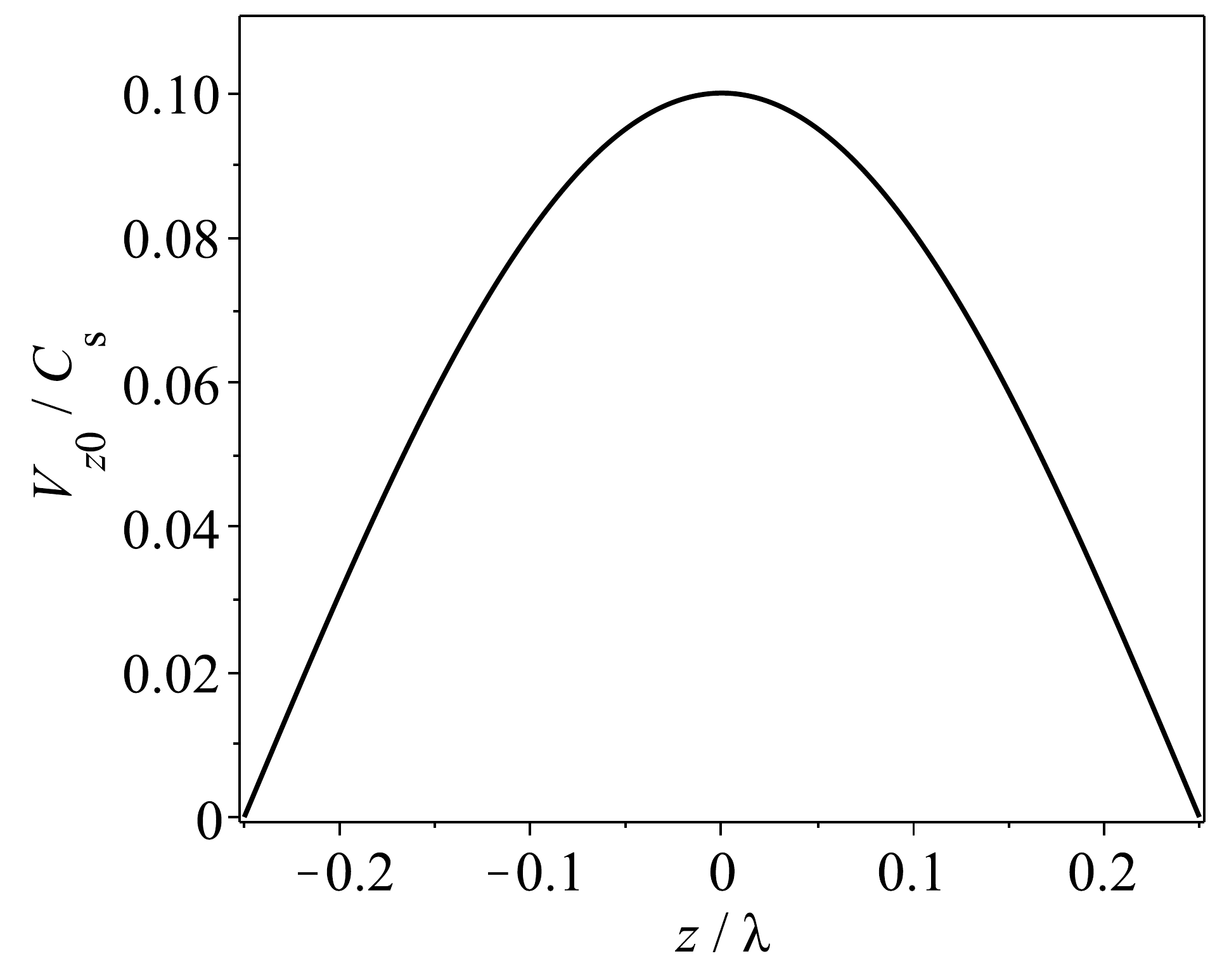}
		\includegraphics[width=0.61\textwidth]{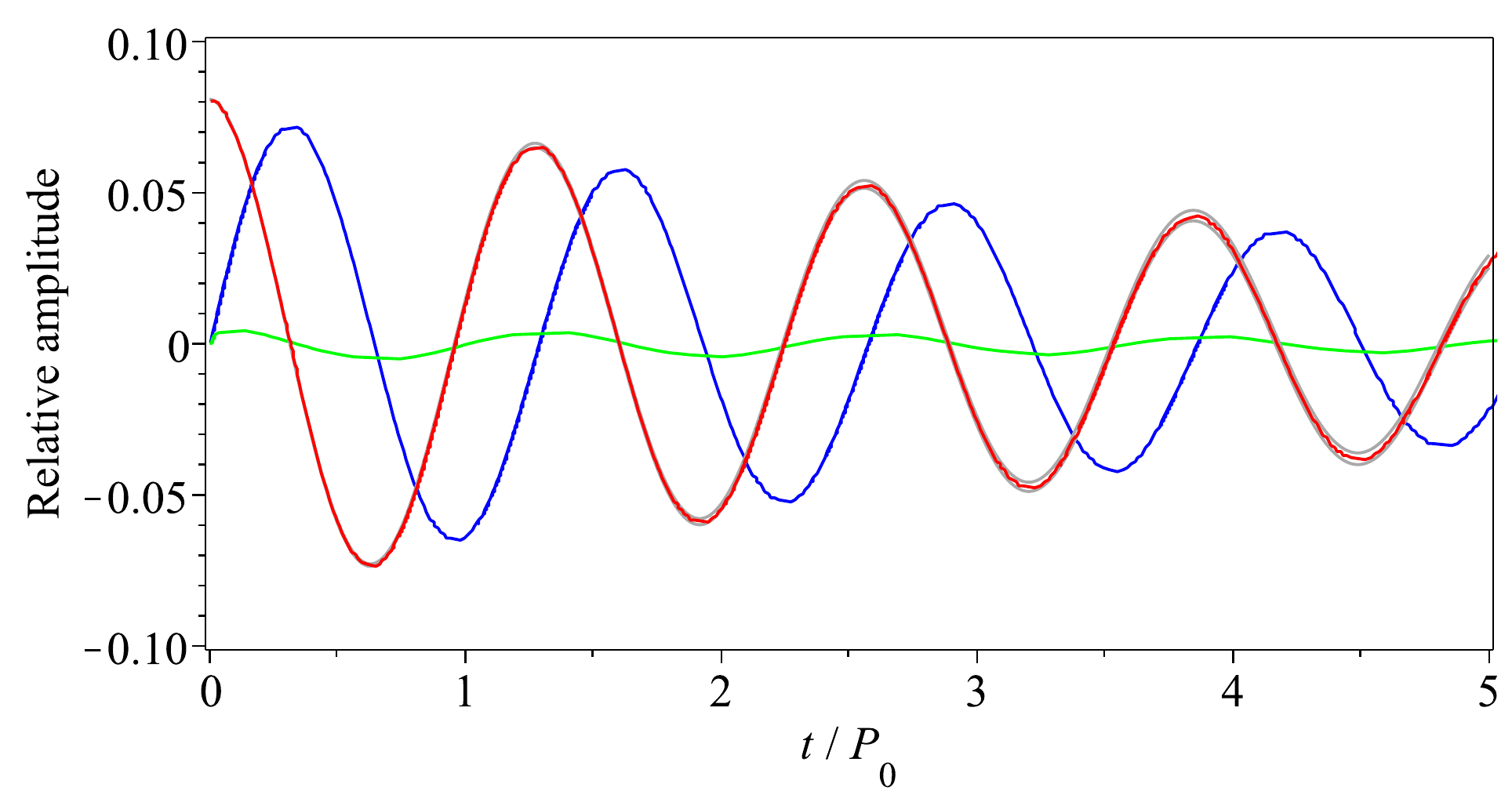}
	\end{center}
	\caption{Left: The form of the initial perturbation of the plasma velocity, $V_{z0} = 0.1 C_\mathrm{s} \cos(2 \pi z/\lambda)$, applied to the loop model described in Sec.~\ref{sec:model}.
		Right: Variations of the slow wave-perturbed plasma velocity, density, and temperature in red, blue, and green, measured at $z=0.1\lambda$ and for $\kappa_\parallel=10\kappa_\mathrm{Sp}$, normalised to $C_\mathrm{s}$, $\rho_0$, and $T_0$, respectively. $C_\mathrm{s} = \sqrt{\gamma k_\mathrm{B}T_0/m}$ is the adiabatic sound speed; $\lambda$ is the wavelength (prescribed by the loop length, $L$, as $\lambda = 2L$); $P_0$ is the adiabatic acoustic oscillation period, $2L/C_s$. {{The grey lines illustrate the numerical error estimate for the velocity perturbation.}}
	}
	\label{fig:1}
\end{figure}

In this work, we solve Eqs.~(\ref{eq_motion})--(\ref{eq:energy}) numerically in the mathematical environment \emph{Maple 2020.2}, {{using the built-in function \textit{pdsolve}. It implements a second order (in space and time) centred, implicit finite difference scheme\footnote{\url{https://www.maplesoft.com/support/help/Maple/view.aspx?path=pdsolve/numeric}}, with timestep 0.02$P_0$ and spacestep 0.02$\lambda$ ($\lambda=2L$) providing the numerical accuracy up to 0.2\% of the equilibrium plasma parameters during the first five oscillation cycles (estimated with the \textit{errorest} keyword of the \textit{pdsolve} command) for the initial perturbation amplitude of 10\%.}}
We do not apply the assumption of {weak conductivity} and vary the field-aligned thermal conduction coefficient $\kappa_\parallel$ from 0.01 to 10 of the standard \emph{Spitzer} value $\kappa_\mathrm{Sp} = 10^{-11}T_0^{5/2}\,\mathrm{W\,m}^{-1}\,\mathrm{K}^{-1}$. {The considered interval of $\kappa_\parallel$ is motivated by previous observational findings. In particular, \citet{2021SSRv..217...34W} (see Sec. 8.1) demonstrated that to account for the coronal polytropic index measured by \citet{2011ApJ...727L..32V},  the thermal conductivity needs to be enhanced by an order of magnitude.}
The following values of the equilibrium plasma parameters are considered: $\rho_0 = 3 \times 10^{-12}$\,kg\,m$^{-3}$ and $T_0 = 6.3$\,MK (both uniform along the loop), $L=180$\,Mm, $m=0.6\times1.67\times10^{-27}$\,kg, typical for coronal loops hosting SUMER oscillations. We excite the fundamental harmonic of a standing slow wave by perturbing the plasma velocity with a harmonic function with maximum at $z=0$ and the wavelength equal to double the loop length, $\lambda=2L$, and apply rigid-wall boundary conditions at $z=\pm 0.25 \lambda$ mimicking the effective slow wave reflection from the transition region and dense chromosphere \citep[e.g.][]{2004A&A...414L..25N}. The form of the initial perturbation and an example of time variations of the plasma velocity, density, and temperature, obtained numerically for e.g. $\kappa_\parallel=10\kappa_\mathrm{Sp}$, are shown in Fig.~\ref{fig:1}.
{All oscillatory signals used in the further analysis are taken at $z=0.1\lambda$.}

\section{Temperature/density phase shifts and amplitudes}
\label{sec:analysis}

\begin{figure}
	\begin{center}
		\includegraphics[width=0.4\textwidth]{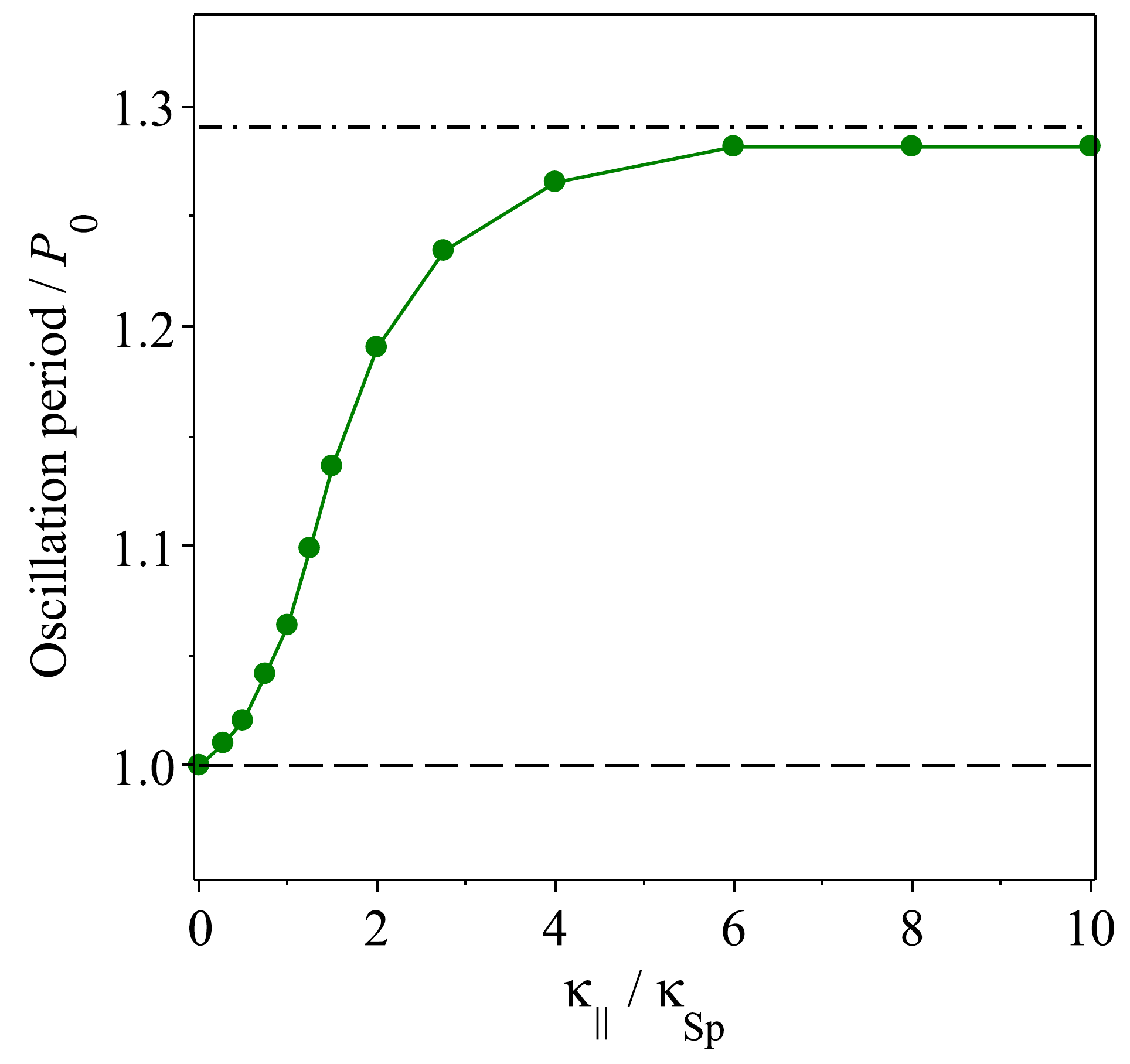}
		\includegraphics[width=0.48\textwidth]{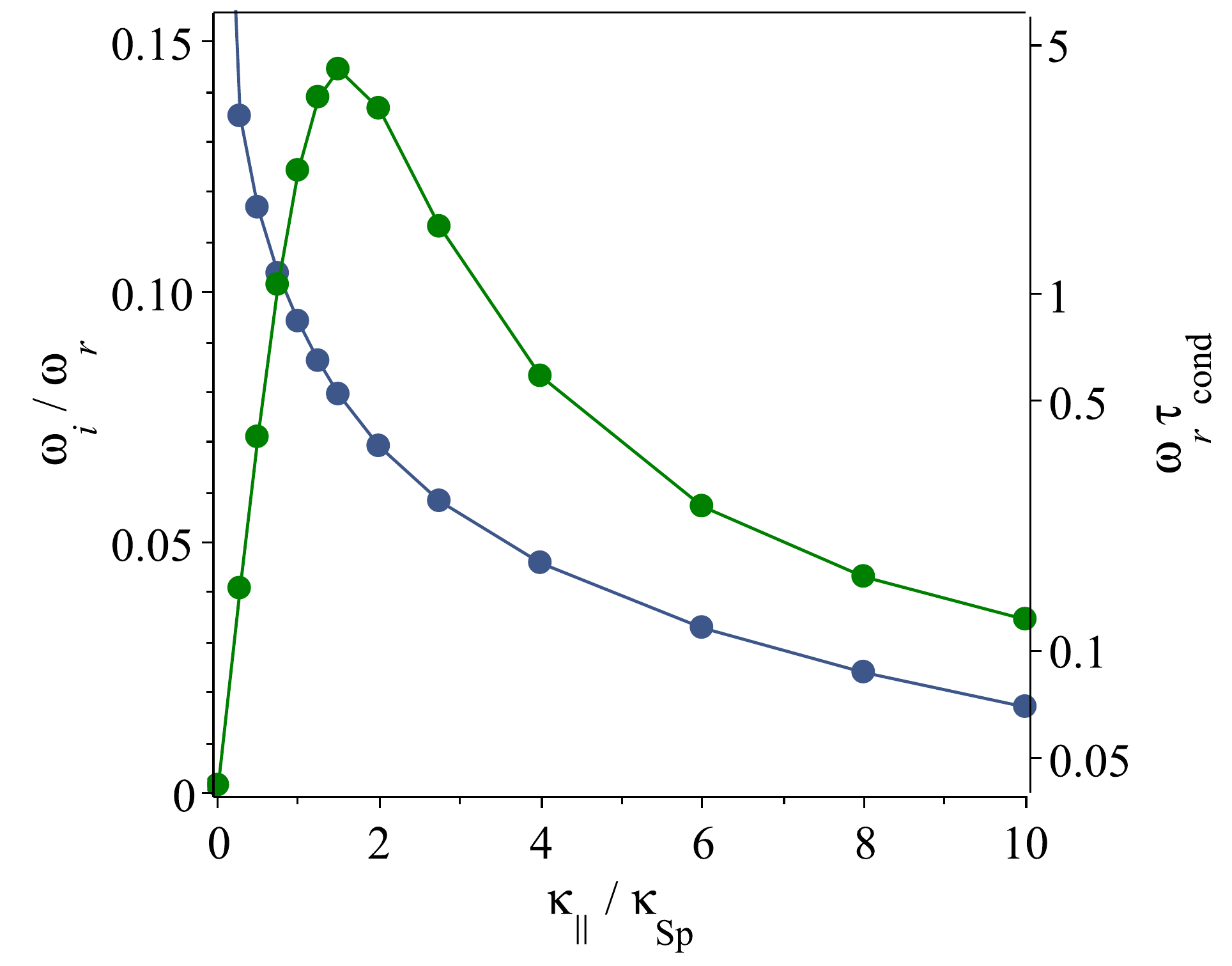}
	\end{center}
	\caption{Left: Slow-wave oscillation period, $P$, estimated empirically via the fast Fourier transform of the numerical solution for the plasma velocity perturbation, vs. the coefficient of the field-aligned thermal conduction normalised to the standard Spitzer value. The horizontal dashed and dot-dashed lines show the period values in the ideal adiabatic  and isothermal limits, obtained with the standard adiabatic and isothermal sound speeds, $C_\mathrm{s}$ and $C_\mathrm{s}/\sqrt{\gamma}$, respectively.
	Right: Dependence of the ratio of the imaginary part $\omega_i=1/\tau_D$ to the real part $\omega_r = 2\pi/P$ of the slow-wave angular frequency $\omega$ on the field-aligned thermal conduction coefficient (green), with $\tau_D$ being the oscillation exponential damping time estimated empirically from the numerical solution.
	{The dark blue line in the right panel shows the dimensionless parameter $\omega_r \tau_\mathrm{cond}$, with $\tau_\mathrm{cond} = \rho_0 C_\mathrm{V}\lambda^2/\kappa_\parallel$ being the characteristic time scale of thermal conduction.}
	}
	\label{fig:2}
\end{figure}

We begin the analysis of the numerically modelled standing slow wave with obtaining the dependence of the oscillation period $P$ on the thermal conduction coefficient $\kappa_\mathrm\parallel$. It is estimated through the fast Fourier transform applied to the plasma velocity perturbations for several different values of $\kappa_\mathrm\parallel$ (see e.g. the red line in Fig.~\ref{fig:1}, right panel) and is presented in the left panel of Fig.~\ref{fig:2}. As one can see, the oscillation period increases by about 30\% with $\kappa_\mathrm\parallel$, from the ideal adiabatic value $P_0$ determined by the standard sound speed $C_\mathrm{s}$, to a new value in the isothermal regime (achieved for high $\kappa_\mathrm\parallel$), determined by the isothermal sound speed $C_\mathrm{s}/\sqrt{\gamma}$. Throughout this work, the loop length $L$ (and therefore the wavelength $\lambda$) is kept constant. In other words, in the {strongly conductive regime}, the effective slow wave speed gets significantly modified, which leads to the corresponding modification in the wave travel time along the loop, i.e. the oscillation period. This empirical result is consistent with previous analytical estimations by e.g. \citet{2021A&A...646A.155D}.

\begin{figure}
	\begin{center}
		\includegraphics[width=0.38\textwidth]{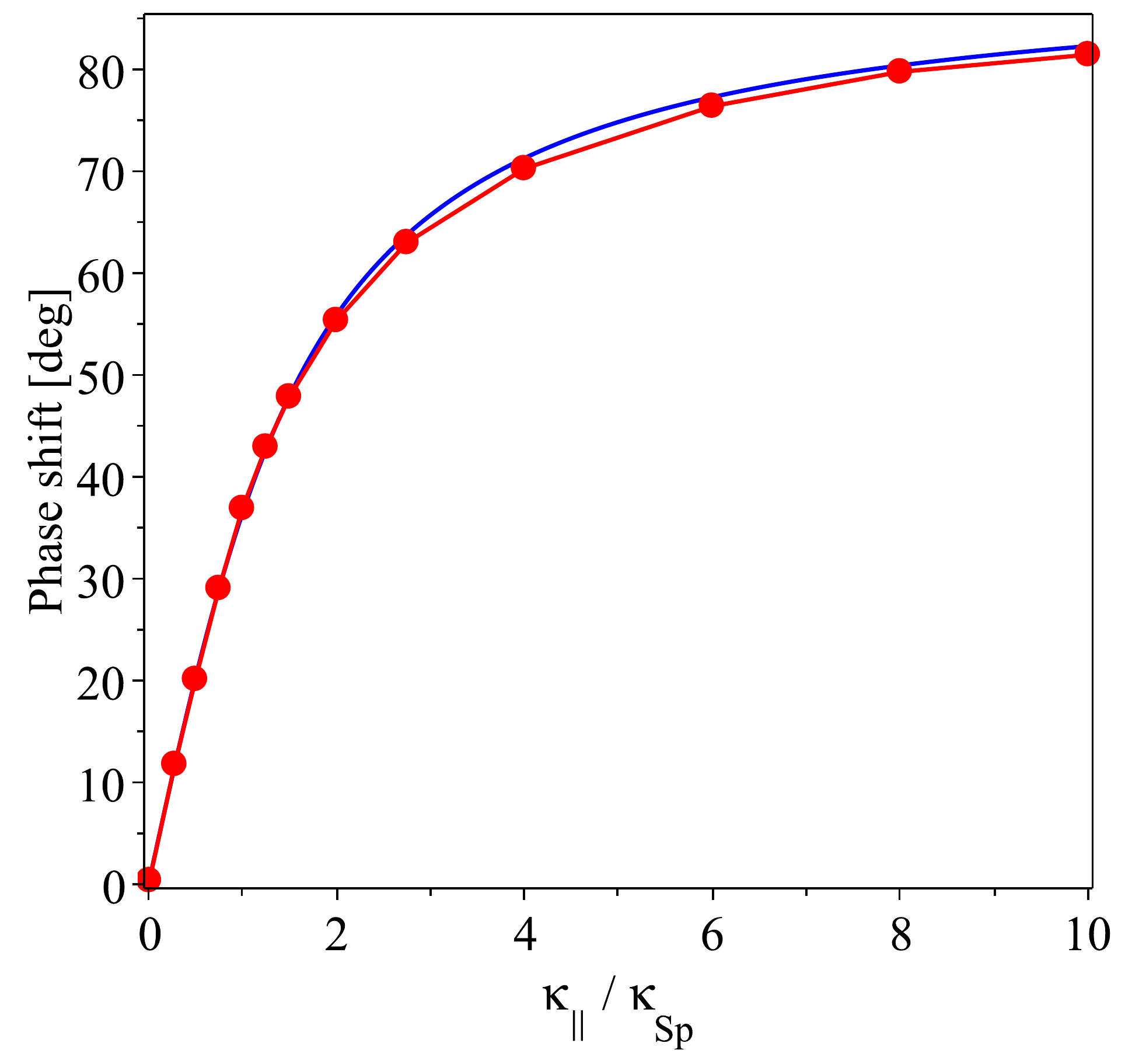}
		\includegraphics[width=0.43\textwidth]{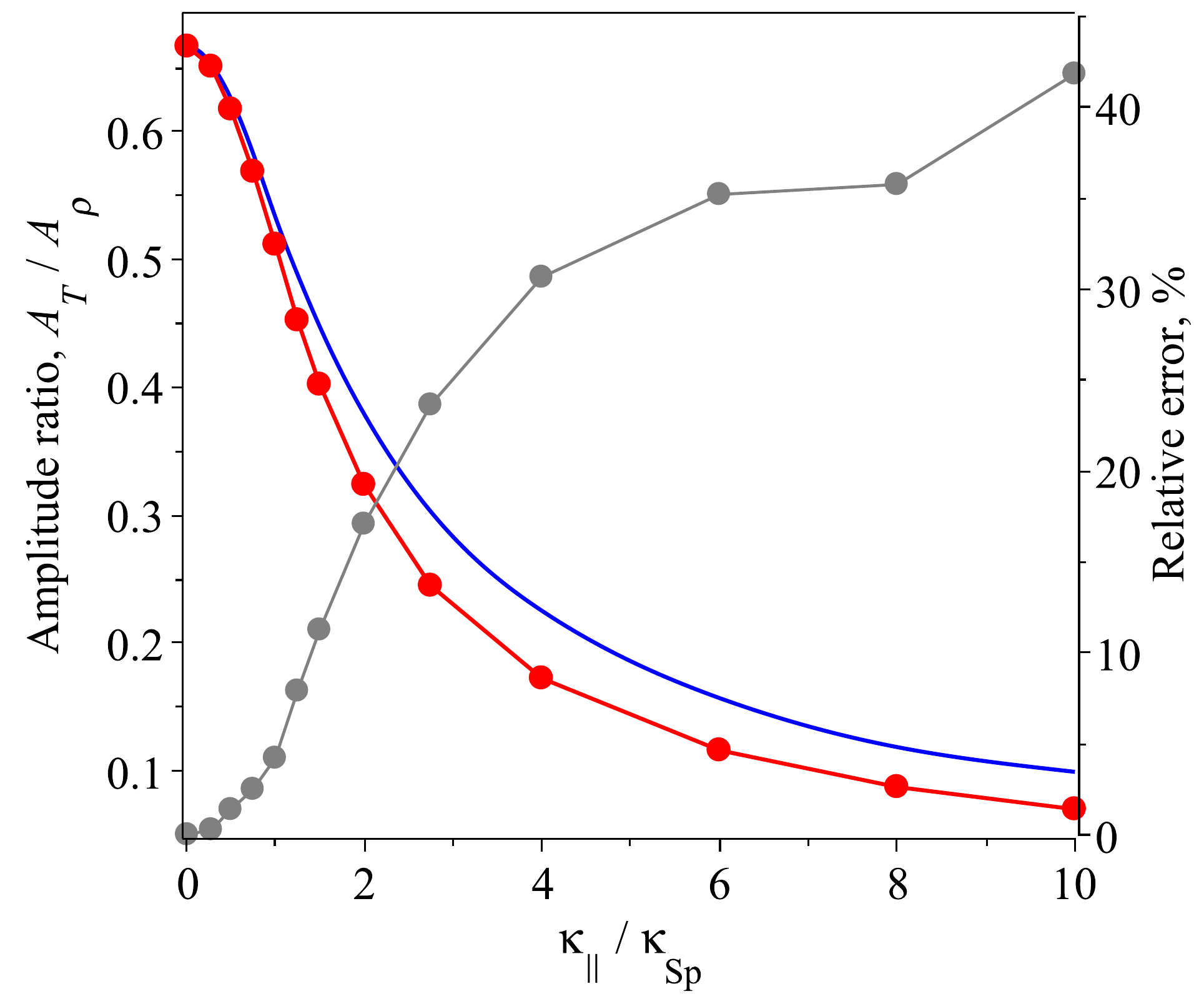}
	\end{center}
	\caption{The phase shift (left) and relative amplitude ratio (right) between plasma temperature and density perturbations by the standing slow wave, obtained from the analysis of the numerical solution of Eqs.~(\ref{eq_motion})--(\ref{eq:energy}) with {full thermal conductivity} as described in Sec.~\ref{sec:analysis} (in red) and from the approximate analytical solutions (\ref{eq:phi_weak})--(\ref{eq:amp_weak}) in a {weakly conductive limit} (in blue). The grey curve in the right panel shows the relative error between the estimations of the temperature and density relative amplitude ratio, shown in red and blue.
	}
	\label{fig:3}
\end{figure}

The phase shifts between density and temperature perturbations are estimated from our numerical solution through the cross-correlation analysis. More specifically, we obtain the time lag $\Delta t$ for which the cross-correlation between density and temperature oscillations is the highest. With this, the phase shift $\Delta \varphi$ is obtained as $\Delta \varphi = (\Delta t/P)\times 360^{\circ}$ for each considered value of $\kappa_\mathrm\parallel$, using the dependence of the oscillation period $P$ on $\kappa_\mathrm\parallel$ obtained above. Thus, the dependence of $\Delta \varphi$ on $\kappa_\mathrm\parallel$, revealed empirically, is shown in the left panel of Fig.~\ref{fig:3} in red. It is seen to vary from $0^{\circ}$ in the ideal adiabatic case to about $80^{\circ}$ in the {strongly conductive regime} with high $\kappa_\mathrm\parallel$ \citep[cf.][]{2022SoPh..297....5P}.

\begin{figure}
	\begin{center}
		\includegraphics[width=0.48\textwidth]{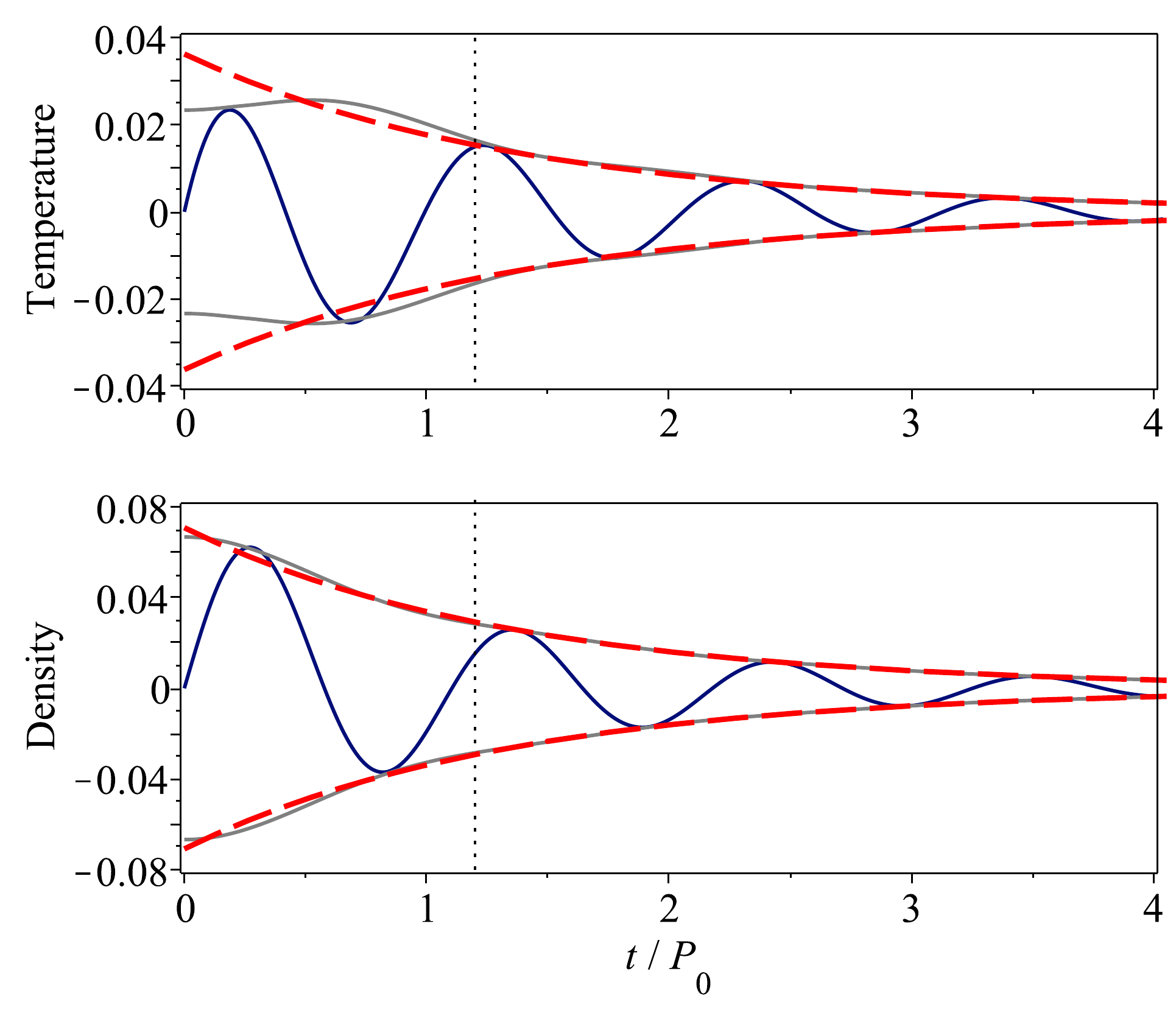}
		\includegraphics[width=0.5\textwidth]{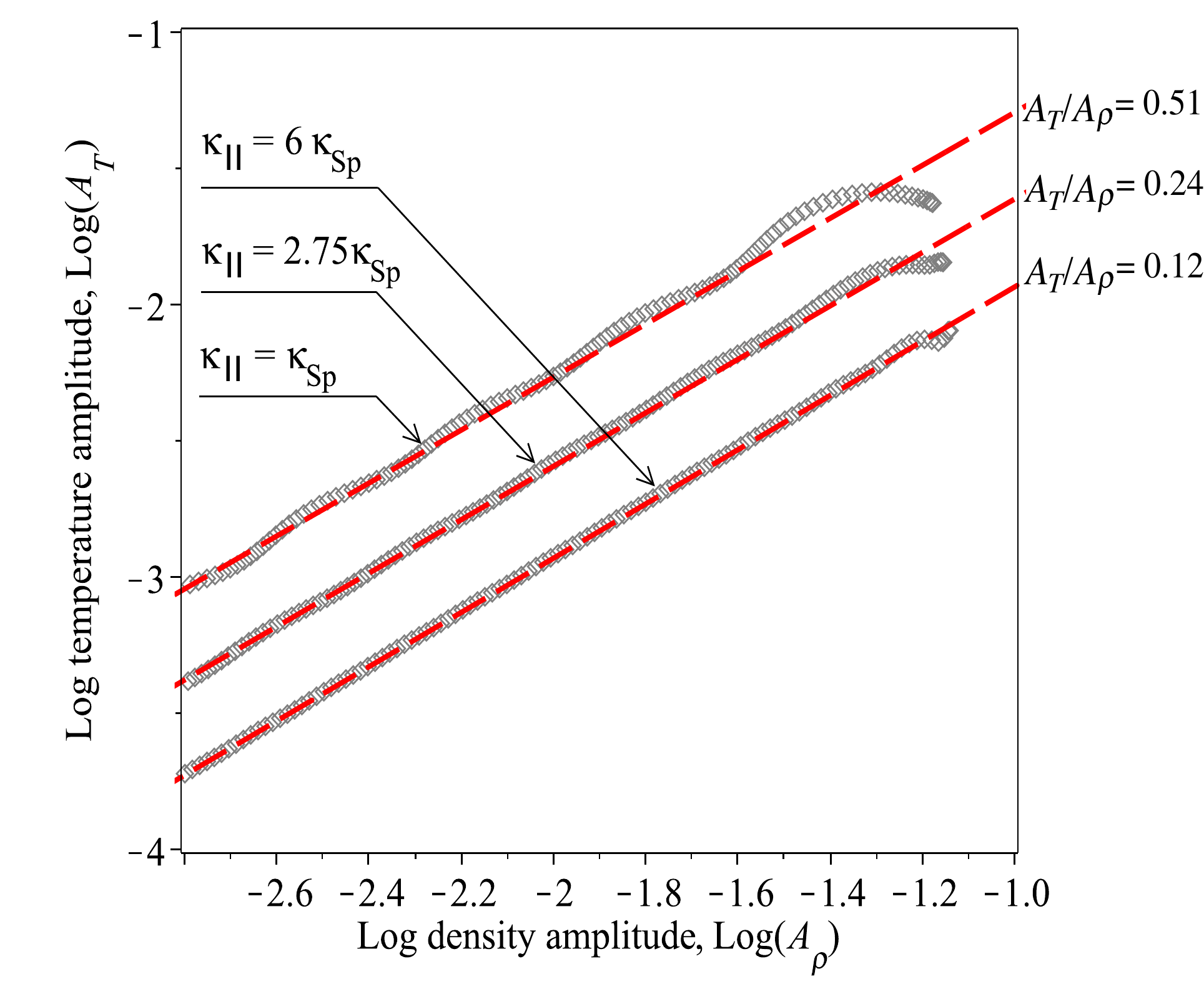}
	\end{center}
	\caption{The time profiles of the temperature and density perturbations by a standing slow wave, obtained numerically from Eqs.~(\ref{eq_motion})--(\ref{eq:energy}) at $z=0.1\lambda$, for $\kappa_\parallel=\kappa_\mathrm{Sp}$, and normalised to $T_0$ and $\rho_0$, respectively (left). The envelopes of the temperature and density perturbations are obtained with the Hilbert transform (grey solid) and by exponential fitting (red dashed). The vertical dotted lines in the left panels indicate the apparent transition time from a non-exponential to exponential damping. The right panel shows examples of the amplitude ratios for several values of $\kappa_\parallel$ (shown in the inlet), estimated with the Hilbert transform (grey diamonds) and by exponential fitting (red dashed). Mind the logarithmic scale in the right panel.
	}
	\label{fig:4}
\end{figure}

For estimating the ratio between temperature and density perturbation amplitudes, we obtain the instantaneous amplitudes $A_T(t)$ and $A_\rho(t)$ as oscillation envelopes by exponential fitting and with the use of the Hilbert transform \citep[see][for apparently the first use of the Hilbert transform for coronal seismology by slow waves]{2019ApJ...884..131R}. The edge effects of the Hilbert transform are mitigated by mirroring the signals with respect to the vertical axis and smoothing the resulting oscillation envelopes over a half of the oscillation period. The examples of $A_T$ and $A_\rho$ for $\kappa_\parallel = \kappa_\mathrm{Sp}$, obtained with the Hilbert transform and their best-fits by decaying exponential functions, are shown in the left panel of Fig.~\ref{fig:4}. It shows, in particular, that actual temperature and density perturbation amplitudes do not obey the exponential law during the first cycle of oscillation, which we attribute to the simultaneous development and rapid decay of the entropy mode \citep{2011A&A...533A..18M, 2021SoPh..296...96Z}.
Thus, associating this mismatch with a possible signature of the slow wave coupling with entropy waves, the ratio $A_T/A_\rho$ in slow waves is estimated via the exponentially decaying instantaneous amplitudes obtained by fitting (the right panel of Fig.~\ref{fig:4}).
{More specifically, the red dashed lines in the right panel of Fig.~\ref{fig:4} show exponentially decaying $A_T(t)$ vs. exponentially decaying $A_\rho(t)$ for three different values of $\kappa_\parallel$. As the slow wave damping rate is the same in perturbations of all plasma parameters, the ratio of exponentially decaying $A_T(t)$ and $A_\rho(t)$ is independent of time. In other words, it may be characterised by the $y$-intercept of the red dashed lines shown in the right panel of Fig. 4 (indeed, if $A_T = \mathrm{const}\times A_\rho$, then $\log(A_T) = \log(\mathrm{const})+\log(A_\rho)$).}
The dependence of the obtained values of $A_T/A_\rho$ on $\kappa_\parallel$ is shown in the right panel of Fig.~\ref{fig:3} in red. It varies from 2/3 ($\equiv \gamma - 1$) in the ideal adiabatic regime to almost zero in the isothermal regime for high $\kappa_\parallel$, in which temperature gradients are effectively smoothed out by thermal conduction. {We also note that the use of a non-exponential damping envelope may lead to underestimated values of $A_T/A_\rho$ in slow waves.}

In addition, this analysis allows us to estimate the conductive damping rate of standing slow waves {and the characteristic thermal conduction time scale $\tau_\mathrm{cond}=\rho_0 C_\mathrm{V}\lambda^2/\kappa_\parallel$} for various values of the conduction coefficient $\kappa_\parallel$ (the right panel in Fig.~\ref{fig:2}). For this, we consider the ratio of the imaginary part of the angular oscillation frequency $\omega_i$ (estimated as the reciprocal of the exponential damping time) to the real part $\omega_r = 2\pi/P$. The obtained non-monotonic behaviour of $\omega_i/\omega_r$ is consistent with the previous analytical estimations by e.g. \citet{2003A&A...408..755D}, with the highest value of about $0.15$, detected for $\kappa_\parallel\approx 2\kappa_\mathrm{Sp}$, being consistent with the observed rapid damping of SUMER oscillations with quality factors (i.e. ratio of the oscillation damping time to period) of about unity \citep[e.g.][]{2019ApJ...874L...1N, 2021SSRv..217...34W}. {The right panel of Fig.~\ref{fig:2} also shows the dimensionless parameter $\omega_r \tau_\mathrm{cond}$ that can be used for a quantitative discrimination between weak and strong conductive limits. Thus, $\omega_r \tau_\mathrm{cond} \gg 1$ and $\omega_i/\omega_r\propto 1/ \omega_r\tau_\mathrm{cond}$ in the weak limit, and $\omega_r \tau_\mathrm{cond} \ll 1$ and $\omega_i/\omega_r\approx\omega_r\tau_\mathrm{cond}$ in the strong limit \citep[see e.g.][]{2014ApJ...789..118K, 2021A&A...646A.155D}. For $\kappa_\parallel\approx 2\kappa_\mathrm{Sp}$ consistent with observations, $\omega_r \tau_\mathrm{cond}$ is about 1 by an order of magnitude, so that neither of those limits is fully justified.}

We now compare the dependences of $\Delta \varphi$ and $A_T/A_\rho$ on $\kappa_\parallel$ obtained from the analysis of our numerical solution to those prescribed by approximate solutions (\ref{eq:phi_weak}) and (\ref{eq:amp_weak}), derived in a {weakly conductive limit} (see the red and blue lines in Fig.~\ref{fig:3}). For both $\Delta \varphi$ and $A_T/A_\rho$, the approximate and numerical solutions seem to perfectly agree for low values of $\kappa_\parallel$, i.e. in the {weakly conductive regime}. For higher $\kappa_\parallel$, the phase shifts $\Delta \varphi$, estimated in the {fully conductive and weakly conductive regimes}, remain well consistent with each other, with a relative error being below a few percent which is practically indistinguishable in observations. In contrast, the amplitude ratio $A_T/A_\rho$ is found to differ significantly from its {weakly conductive estimation} for $\kappa_\parallel \gtrsim \kappa_\mathrm{Sp}$. The relative error of this offset is seen to be about 5\% for $\kappa_\parallel = \kappa_\mathrm{Sp}$ and reaches 30--40\% for higher $\kappa_\parallel$.

\section{Effective adiabatic index}
\label{sec:gamma}

In this section, we demonstrate the application of the obtained wave parameters to probing the effective adiabatic index of the coronal plasma, $\gamma_\mathrm{eff}$, and assess the suitability of a commonly used polytropic assumption for it.

\begin{figure}
	\begin{center}
		\includegraphics[width=0.4\textwidth]{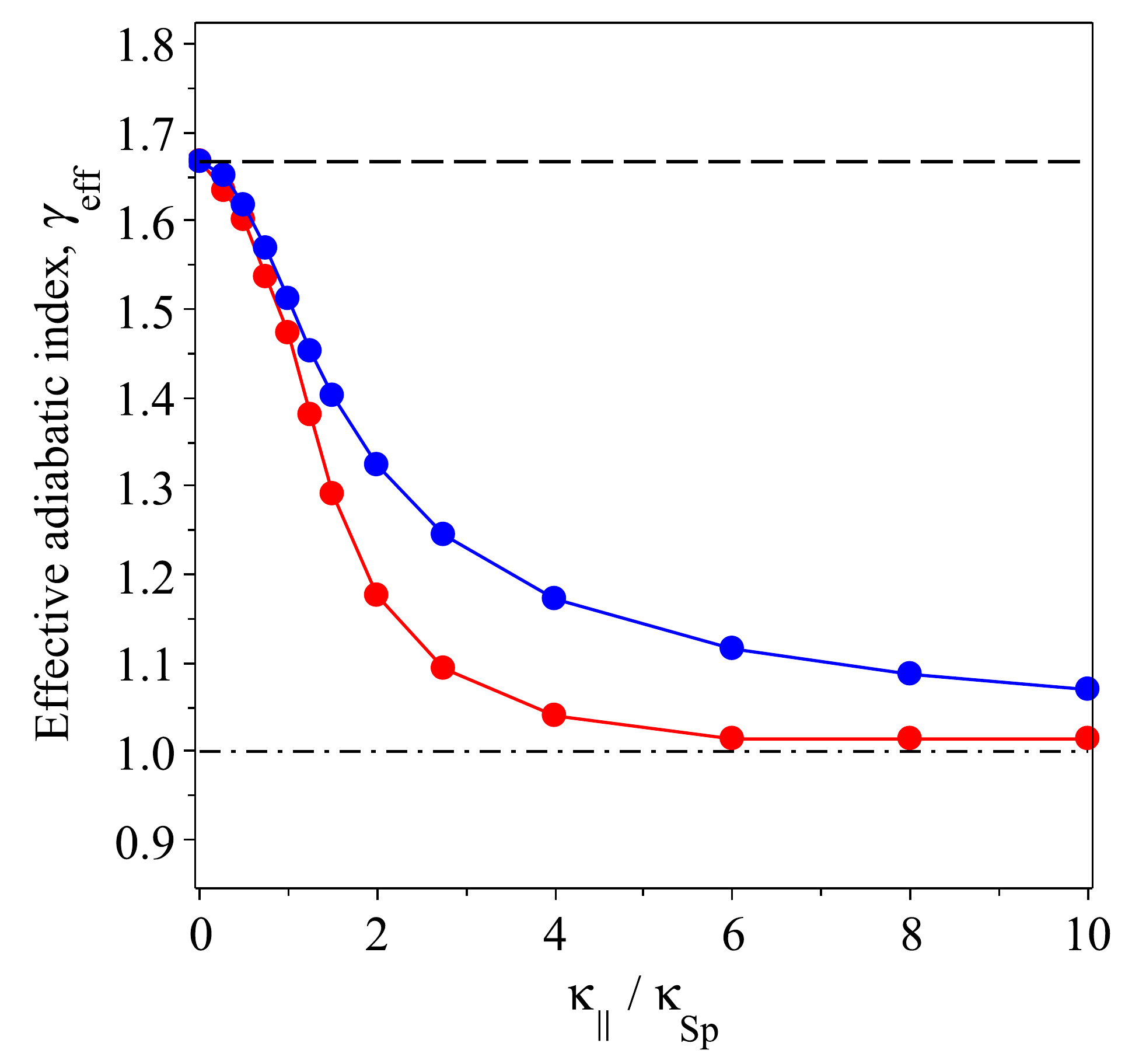}
		\includegraphics[width=0.4\textwidth]{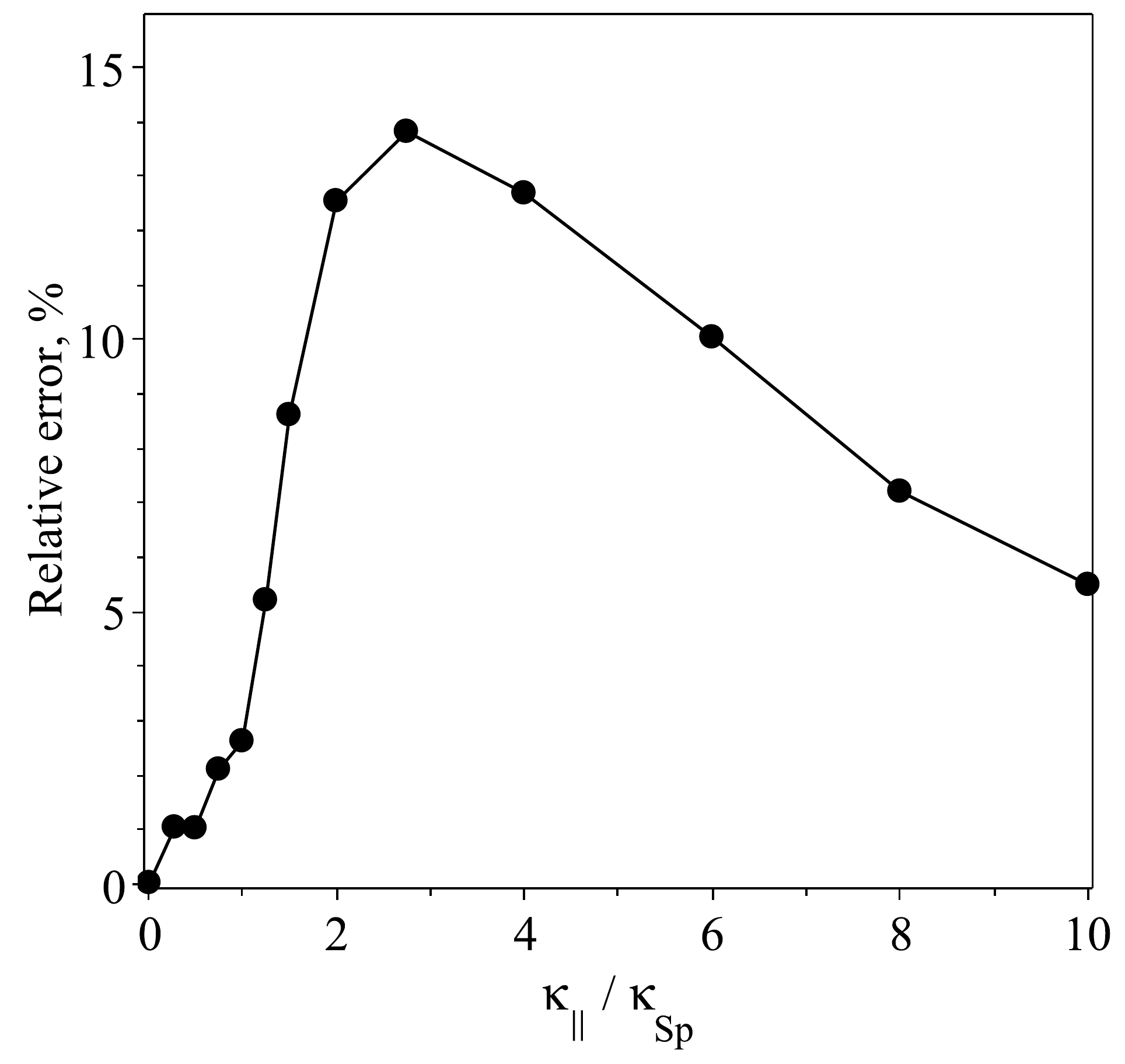}
	\end{center}
	\caption{The dependence of the effective adiabatic index $\gamma_\mathrm{eff}$ of the coronal plasma on the field-aligned thermal conduction coefficient $\kappa_\parallel$ (left), estimated numerically as ratio of the effective slow wave speed to the standard sound speed (red), see Eqs.~(\ref{eq:gamm_speeds}) and (\ref{eq:gamm_periods}), and under the polytropic assumption (blue) with Eq.~(\ref{eq:gamma_poly}). The horizontal dashed and dot-dashed lines indicate the values of $\gamma_\mathrm{eff}$ in the ideal adiabatic case (5/3) for low $\kappa_\parallel$ and in the isothermal regime (1) for high $\kappa_\parallel$, respectively.
	The right panel shows the relative error between the estimations of $\gamma_\mathrm{eff}$, shown in red and blue in the left panel.
	}
	\label{fig:5}
\end{figure}

Following e.g. \citet{2018ApJ...860..107W} and \citet{2019PhPl...26h2113Z}, we define $\gamma_\mathrm{eff}$ as a measure of the deviation of the observed phase speed $V_\mathrm{p}$ of slow waves affected by non-adiabatic effects (the field-aligned thermal conduction in our model) from the standard sound speed $C_\mathrm{s}$, i.e.
\begin{equation}
	\gamma_\mathrm{eff} = \gamma \left(\frac{V_\mathrm{p}}{C_\mathrm{s}} \right)^2.\label{eq:gamm_speeds}
\end{equation}
In the solar corona, the standard sound speed can be estimated as $C_\mathrm{s} \mathrm{[km/s]} \approx 152 \sqrt{T_0\mathrm{[MK]}}$. On the other hand, as the wavelength of the discussed standing slow wave is prescribed by the loop length and thus remains constant, we can use the dependence of the oscillation period $P$ on $\kappa_\mathrm\parallel$ obtained in Sec.~\ref{sec:analysis} (see Fig.~\ref{fig:2}) as a proxy (observable parameter) of the slow wave phase speed $V_\mathrm{p}$. With this, the definition of $\gamma_\mathrm{eff}$ (\ref{eq:gamm_speeds}) can be re-written as
\begin{equation}
	\gamma_\mathrm{eff} = \gamma \left(\frac{P_0}{P} \right)^2,\label{eq:gamm_periods}
\end{equation}
with $P_0$ being the slow wave oscillation period in the ideal adiabatic case. The dependence of $\gamma_\mathrm{eff}$ estimated by Eq.~(\ref{eq:gamm_periods}) on the field-aligned thermal conduction coefficient $\kappa_\parallel$ is shown in the left panel of Fig.~\ref{fig:5} in red, using the dependence of the oscillation period $P$ on $\kappa_\parallel$ shown in Fig.~\ref{fig:2}. As expected, the obtained values of $\gamma_\mathrm{eff}$ decrease with $\kappa_\parallel$ from 5/3 to 1 in the ideal adiabatic and isothermal regimes, respectively.

In the polytropic assumption, i.e. assuming the plasma density and pressure perturbations to be connected through a power-law as $p\propto\rho^{\gamma_\mathrm{eff}}$, $\gamma_\mathrm{eff}$ can be estimated through the ratio of the instantaneous relative amplitudes $A_T$ and $A_\rho$ of temperature and density perturbations \citep[][]{2011ApJ...727L..32V} as
\begin{equation}
	\gamma_\mathrm{eff} \approx \frac{A_T}{A_\rho} + 1.\label{eq:gamma_poly}
\end{equation}
Despite being not strictly consistent with the observed non-zero phase difference between temperature and density perturbations, this assumption is widely used for probing the effective adiabatic index of the corona with both standing \citep[e.g.][]{2015ApJ...811L..13W, 2019ApJ...884..131R} and propagating \citep[e.g.][]{2011ApJ...727L..32V, 2018ApJ...868..149K} slow waves.

The dependence of $\gamma_\mathrm{eff}$ (\ref{eq:gamma_poly}) on $\kappa_\parallel$, using $A_T/A_\rho$ estimated empirically in Sec.~\ref{sec:analysis} (the red line in Fig.~\ref{fig:3}, right panel), is shown in the left panel on Fig.~\ref{fig:5} in blue. Its comparison with $\gamma_\mathrm{eff}$ (\ref{eq:gamm_periods}), as ratio of the effective wave speed to the standard sound speed (i.e. {ratio of the slow wave period in adiabatic case to the observed period}), justifies the use of the polytropic assumption in a {weakly conductive regime} (for $\kappa_\parallel \lesssim \kappa_\mathrm{Sp}$ and $\gamma_\mathrm{eff}$ being between approximately 1.5 and 5/3) and reveals the relative errors (the right panel of Fig.~\ref{fig:5}) comparable to those detected in observations for $\kappa_\parallel > \kappa_\mathrm{Sp}$, that reach the maximum of 14\% for $\kappa_\parallel \approx 3\kappa_\mathrm{Sp}$. Even in the isothermal regime with high $\kappa_\parallel$, the mismatch between $\gamma_\mathrm{eff}$ (\ref{eq:gamm_periods}) and its polytropic approximation (\ref{eq:gamma_poly}) remains above 5\% {{(e.g. for $\kappa_\parallel = 6 \kappa_\mathrm{Sp}$, $\gamma_\mathrm{eff}$ is about 1.0 by Eq.~(\ref{eq:gamm_periods}) and is about 1.1 by Eq.~(\ref{eq:gamma_poly})).}}

\section{Discussion and conclusions}
\label{sec:disc}

The applicability of a weakly non-adiabatic because of finite thermal conduction along the field and polytropic assumptions to coronal seismology with slow waves has been studied in this work. We numerically modelled a 1D evolution of the fundamental harmonic of a standing slow wave in a strongly magnetised coronal plasma loop, with the field-aligned thermal conduction as the dominant wave damping mechanism and the conduction coefficient $\kappa_\parallel$ as a free parameter. In the model, no restrictions on the effectiveness of thermal conduction were imposed. The time profiles of the plasma velocity, density, and temperature perturbations have been treated as effective observables to which the standard data analysis techniques, such as the fast Fourier transform and cross-correlation analysis, and more advanced Hilbert transform, were applied. The outcomes of this analysis have been compared to the approximate analytical solutions. The main results of this work can be summarised as:
\begin{itemize}
	\item The finite thermal conductivity along the field modifies the effective speed of slow waves, which leads to the modification of the observed oscillation period by up to 30\% from the value estimated in the ideal regime and used in the {weakly conductive limit}. Accounting for additional non-adiabatic effects, such as e.g. the wave-induced misbalance between coronal heating and cooling processes \citep{2021PPCF...63l4008K}, may make this modification even stronger.
	
	\item The dependences of the phase shift $\Delta \varphi$ between the loop's temperature and density perturbations on the thermal conductivity $\kappa_\parallel$, estimated in the {strongly and weakly conductive cases}, are well consistent with each other for both low and high values of $\kappa_\parallel$. The obtained ratio of temperature and density relative amplitudes $A_T/A_\rho$, in contrast, agrees with the {weak conduction theory} for $\kappa_\parallel \lesssim \kappa_\mathrm{Sp}$ only. For higher $\kappa_\parallel$, the mismatch can reach up to 30--40\%, which clearly requires accounting for higher-order non-adiabatic effects.
	
	\item From the practical point of view, the previous finding allows one to reduce the analytical solutions for $\Delta \varphi$ and $A_T/A_\rho$ obtained with {full conductivity} \citep[see e.g. Eqs.~(51) and (52) in][]{2021SSRv..217...34W}, which are essentially coupled through two unknowns $\kappa_\parallel$ and $\gamma_\mathrm{eff}$ and therefore cannot be used independently, to Eq.~(\ref{eq:phi_weak}) for $\Delta \varphi$ and
	\begin{equation}
		\frac{A_T}{A_\rho}=\frac{(\gamma-1)\cos \Delta \varphi}{1-2\pi\gamma d \chi (\gamma/\gamma_\mathrm{eff})},\label{eq:amp_strong}
	\end{equation}
	where $d \equiv {(\gamma-1)\kappa_\parallel m}/{\gamma k_\mathrm{B}C_\mathrm{s}^2 P_0 \rho_0}$, $\chi \equiv \omega_i/\omega_r$, and $\gamma_\mathrm{eff}$ is given by Eq.~(\ref{eq:gamm_speeds}). As such, Eqs.~(\ref{eq:phi_weak}) and (\ref{eq:amp_strong}) are de-coupled with respect to $\kappa_\parallel$ and $\gamma_\mathrm{eff}$ (the right-hand side of Eq.~(\ref{eq:phi_weak}) has $\kappa_\parallel$ only), which would significantly simplify their future seismological applications without the loss of accuracy.
	
	\item The polytropic assumption (\ref{eq:gamma_poly}) can be used for probing the effective adiabatic index of the coronal plasma, $\gamma_\mathrm{eff}$, in the {weakly conductive regime} only, i.e. with $\kappa_\parallel \lesssim \kappa_\mathrm{Sp}$ and small deviations of $\gamma_\mathrm{eff}$ from the adiabatic value 5/3. For $\kappa_\parallel > \kappa_\mathrm{Sp}$ or if $\gamma_\mathrm{eff}$ is deemed to differ from 5/3 by more than 10\%, it should be estimated either as a ratio of the observed slow wave oscillation period (phase speed) to the period expected in the ideal adiabatic case (standard sound speed) or via the ratio of relative amplitudes $A_T/A_\rho$ using Eq.~(\ref{eq:amp_strong}). Otherwise, the relative errors may reach up to 14\% \citep[cf. 7\% uncertainty in the estimation of $\gamma_\mathrm{eff}$, detected by][for example]{2018ApJ...868..149K}.
	
	\item As an additional side result of this work, a non-exponential damping of slow waves during approximately the first cycle of oscillation was detected with the use of the Hilbert transform. Similarly to the transition time from a Gaussian to exponential damping of coronal kink oscillations {by mode coupling with torsional Alfv\'en waves} \citep[e.g.][]{2017A&A...600A..78P}, the revealed non-exponential damping of slow waves can be used as an indirect signature of the entropy mode evolution with yet unexploited seismological potential. {In particular, this non-exponential damping of slow waves is seen to be more pronounced in the perturbation of plasma temperature and for lower values of $\kappa_\parallel$ in our analysis, the reason for which is to be understood.}
\end{itemize}

This work establishes an important ground for the application of the method of coronal seismology by slow waves in strongly non-adiabatic conditions. Moreover, the performed analysis can be readily generalised for additional non-adiabatic effects, such as compressive viscosity, optically thin radiation and enigmatic coronal heating, and used for validation of the corresponding theories \citep[e.g.][]{2022SoPh..297....5P} without the need to deploy full-scale viscous 3D MHD simulations \citep[e.g.][]{2022ApJ...926...64O} or dedicated MHD spectral codes \citep[e.g.][]{2020ApJS..251...25C}.

\section*{Conflict of Interest Statement}

The authors declare that the research was conducted in the absence of any commercial or financial relationships that could be construed as a potential conflict of interest.

\section*{Author Contributions}

DYK is the sole author and is responsible for the entire content of this work.

\section*{Funding}
The work was supported by the STFC Consolidated Grant ST/T000252/1.

\section*{Acknowledgments}
The author is grateful to Dr Dmitrii Zavershinskii, Prof Valery Nakariakov, and members of the International Online Team \lq\lq Effects of Coronal Heating/Cooling on MHD Waves\rq\rq\ for inspiring discussions and comments.


\section*{Data Availability Statement}
The original contributions presented in the study are included in the article, further inquiries can be directed to the corresponding author.

\bibliographystyle{Frontiers-Harvard} 



\end{document}